\documentclass[prb,twocolumn,aps,floatfix]{revtex4}

\usepackage{graphicx}
\usepackage{bm}
\usepackage{amssymb}
\usepackage{amsmath}
\usepackage{subfigure}
\usepackage{tikz}
\usetikzlibrary{shapes,arrows}
\usepackage{amsmath}
\usepackage{appendix}
\usepackage{multirow}
\usepackage{booktabs}

\newcommand{\tabf}{\hspace*{0.2em}}

\begin{document}

\title{Lattice Green's functions for kagome, diced and hyperkagome lattices}
\author{Vipin Kerala Varma and Hartmut Monien}
\affiliation{Bethe Center for Theoretical Physics, Universit\"{a}t Bonn, Germany\\}
\date{\today}
\begin{abstract}
Lattice Green's functions (LGF) and density of states (DOS) for non-interacting models on 3 related lattices are presented. The DOS and LGF at the origin for the kagome and diced lattices are rederived. Furthermore, from the form obtained for the DOS of the hyperkagome lattice, it seems highly suggestive that a closed form expression can be obtained if a particular complicated yet highly symmetrical integral can be solved. The results have been checked numerically. The continuum of energy states in the hyperkagome lattice and the appearance of a zero in the kagome lattice may be explicitly seen from the formulae obtained.
\end{abstract}

\maketitle

Lattice Green's functions (LGF) of systems are ubiquitous\cite{Katsura, Economou} in solid state physics appearing in problems of lattice vibrations, of spin wave theory of magnetic systems, of localized oscillation modes at lattice defects, and in combinatorial problems in lattices\cite{Guttmann}. The LGF $G(t,\vec{\mathbf{r}})$ in the complex energy variable $t = s - i\epsilon$ between the (arbitrary) lattice origin and the point $\vec{\mathbf{r}}$ is defined as the solution to the inhomogenous difference equation
\begin{equation}
 [2t - H(\vec{\mathbf{r}})]G(t,\vec{\mathbf{r}}) = 2\delta_{\vec{\mathbf{r}},\vec{\mathbf{0}}}
\end{equation}
corresponding to the linear, Hermitian, time-independent operator $H(\vec{\mathbf{r}})$. This operator is generally taken to be the discrete Laplacian for lattice problems. Its action on an $n$-variable function $\phi(\vec{\mathbf{r}} \equiv \{r_1, r_2 \cdots r_n\})$ is given by
\begin{equation}
 H(\vec{\mathbf{r}})\phi(\vec{\mathbf{r}}) = \sum_{\vec{\mathbf{\Delta}} \in \mathcal{N}}\phi(\vec{\mathbf{r}} + \vec{\mathbf{\Delta}}) - z\phi(\vec{\mathbf{r}}),
\end{equation}
where $\mathcal{N}$ denotes the set of nearest neighbours of $\vec{\mathbf{r}}$ and $z \equiv |\mathcal{N}|$ is the number of such elements. In subsequent analysis, we will neglect the second term in Eq. (2) because it merely adds a constant shift to the energy variable $t$. With that Eq. (1) defining the LGF is expressed as
\begin{equation}
 [2tG(t,\vec{\mathbf{r}}) - \sum_{\vec{\mathbf{\Delta}} \in \mathcal{N}}G(t,\vec{\mathbf{r}} + \vec{\mathbf{\Delta}})] = 2\delta_{\vec{\mathbf{r}},\vec{\mathbf{0}}}.
\end{equation}
The above equation may be readily solved by taking its Fourier transform into $\vec{\mathbf{k}}$-space, noting that $G(t,\vec{\mathbf{r}} + \vec{\mathbf{\Delta}}) \leftrightarrow \textrm{e}^{(\vec{\mathbf{k}}.\vec{\mathbf{\Delta}})}\tilde{G}(t,\vec{\mathbf{k}})$, where $\tilde{G}$ denotes the Fourier transformed variable. For a simple cubic lattice in $d$-dimensions with unit lattice distances, this can be written, after inverse Fourier transforming, as
\begin{equation}
 G(t,\vec{\mathbf{r}}) = (\frac{1}{2\pi})^d\idotsint_{-\pi}^{\pi}\frac{\prod_{i=1}^d\mathrm{d}k_i\textrm{e}^{\vec{\mathbf{k}}.\vec{\mathbf{r}}}}{t - \omega(\vec{\mathbf{k}})}.
\end{equation}
where $\omega(\vec{\mathbf{k}}) = \sum_i^d \cos{(k_i)}$. The general form for the LGF for other lattices like body-centred cubic, face-centred cubic remain the same\cite{Katsura}. In many cases such integrals may be exactly solved in terms of the elliptic integrals of the first and second kind. Once a solution for the LGF is obtained, we may obtain information about the eigenvalues and eigenfunctions of the operator $H(\vec{\mathbf{r}})$. And any solution of the inhomogenous equation may be constructed using $G(t,\vec{\mathbf{r}})$. An important quantity of the lattice problem is the density of states (DOS) $\rho(s)$, which will be a focus of this paper, given as
\begin{equation}
 \rho(s) = \lim_{\epsilon \to +0}\frac{1}{\pi}\Im{[G(t,\vec{\mathbf{0}})]},
\end{equation}
where $\Im$ refers to the imaginary part.\\
In this brief report, we present results for three related lattices at $\vec{\mathbf{r}} = \vec{\mathbf{0}}$. In the first section, we derive the LGF and DOS for the kagome lattice, which is shown in Fig. 1(a). In the second section we perform similar calculations for the diced lattice, which is shown in Fig. 1(b). Some of these results were obtained earlier\cite{HoriguchiChen} but we repeat it here to correct a minor error in Ref. [4]. And in the final section, we attempt to obtain a closed form solution for LGF and DOS for the hyperkagome lattice, which is shown in Fig. 2(a). We could not solve this exactly in a closed form but numerical evaluation of this integral to obtain the DOS does, however, compare well with a previously published purely numerical evaluation of the DOS.

\section{Kagome lattice}
The kagome lattice has 3 atoms per unit cell labelled $A, B, C$ in Fig. 1(a). The side-length of the simple triangle $a = 1$ and the vertical height $b = \sqrt3/2$. We use the general formalism presented before to calculate the LGF at the origin using Eq. (3) and the DOS from the imaginary part using Eq. (5). The difference here is that we need to consider LGFs at and between the different sublattices $\alpha, \beta$ denoted as $G^{\alpha \beta}$. Then the right hand side of Eq. (3) is multiplied by an additional Kronecker delta factor $\delta_{\alpha,\beta}$ and Eq. (5) is modified to
\begin{equation}
  \rho(s) = \lim_{\epsilon \to +0}\frac{1}{N\pi}\Im{[\sum_{\alpha}G^{\alpha \alpha}(t,\vec{\mathbf{0}})]},
\end{equation}
where $N$ denotes the number of sublattices.\\
Taking the Fourier transform of Eq. (3) with respect to $\vec{\mathbf{r}}$, the lattice Green's functions, evaluated at the origin, for the $\nu$-sublattice with respect to the other 3 sublattices may be written as
\begin{eqnarray}
&t\tilde{G}^{A\nu}-\tilde{G}^{B\nu}\cos{(ak_x + bk_y)}-\tilde{G}^{C\nu}\cos{(ak_x - bk_y)} = \delta_{A\nu}, \nonumber \\
&-\tilde{G}^{A\nu}\cos{(ak_x + bk_y)}+t\tilde{G}^{B\nu}-\tilde{G}^{C\nu}\cos{(2ak_x)} = \delta_{B\nu} \nonumber \\
&-\tilde{G}^{A\nu}\cos{(ak_x - bk_y)}-\tilde{G}^{B\nu}\cos{(2ak_x)}+t\tilde{G}^{C\nu} = \delta_{C\nu}, \nonumber \\ \raggedleft
\end{eqnarray}
where $\nu = A, B, C$ as before. The LGFs $\tilde{G}^{\alpha \beta}$ depend on $(t, k_x, k_y)$ but has been suppressed in the above equation for notational simplicity. It is a straightforward task to solve for $\tilde{G}^{\alpha \beta}$ from the above set of linear equations by setting particular values for $\nu$. Then we may write the ``diagonal'' lattice Green functions i.e. $\alpha = \beta$ as
\begin{eqnarray}
& &\tilde{G}^{AA} = [t^2 - \cos^{2}{(2ak_x)}]/\mathcal{D}, \nonumber \\
& &\tilde{G}^{BB} = [t^2 - \cos^{2}{(ak_x - bk_y)}]/\mathcal{D}, \nonumber \\
& &\tilde{G}^{CC} = [t^2 - \cos^{2}{(ak_x + bk_y)}]/\mathcal{D}, \nonumber \\ \raggedleft
\end{eqnarray}
where $\mathcal{D} \equiv (t+1)[t(t-1) - \cos{(2ak_x)}(\cos{(2ak_x)} + \cos{(2bk_y)})]$. The ``off-diagonal'' lattice Green functions may be similarly calculated as well from Eq. (7).

From the knowledge of the Fourier-space diagonal LGFs $\tilde{G}^{\alpha \alpha}$ in Eq. (8), the DOS $\rho_{\textrm{kag.}}(s)$, as a function of the real energy variable $s = \Re{[t]}$, for the kagome lattice may be calculated using Eq. (6)
\begin{eqnarray}
\rho_{\textrm{kag.}}(s) = \frac{1}{3\pi}\Im{\textrm{\huge{\{}}\frac{1}{t+1} + \mathcal{F}^{-1}[2\frac{2t-1}{\tau ' - \omega_k '}]\textrm{\huge{\}}}},
\end{eqnarray}
where $\tau ' \equiv 2s^2 - 2s - 1$, $\omega_k ' \equiv \cos{(4ak_x)} + 2\cos{(2ak_x)}\cos{(2bk_y)}$ and $\mathcal{F}$ denotes the Fourier transform of the $\vec{\mathbf{k}}$ variable within the limits $\pm \pi/a$, $\pm \pi/b$. That is,
\begin{equation}
 \rho_{\textrm{kag.}}(s) = \frac{\delta(s+1)}{3} + \frac{2}{3}|2s - 1|\rho_{\textrm{tri}}(\tau ').
\end{equation}
It may be checked that the spectral density integrates to one i.e. $\int_{-1}^{2}\rho_{\textrm{kag.}}(s)\,\mathrm{d}s = 1$. The expression in Eq. (10) checks with that in Ref. [5] which provided no derivation of the result. We additionally have obtained the diagonal LGFs, in Fourier space, for the kagome lattice in Eq. (8).
\section{Diced lattice}
\begin{figure}[ttp]
\centering
\subfigure[]{\label{dicedlattice}
\includegraphics*[width=4.4cm]{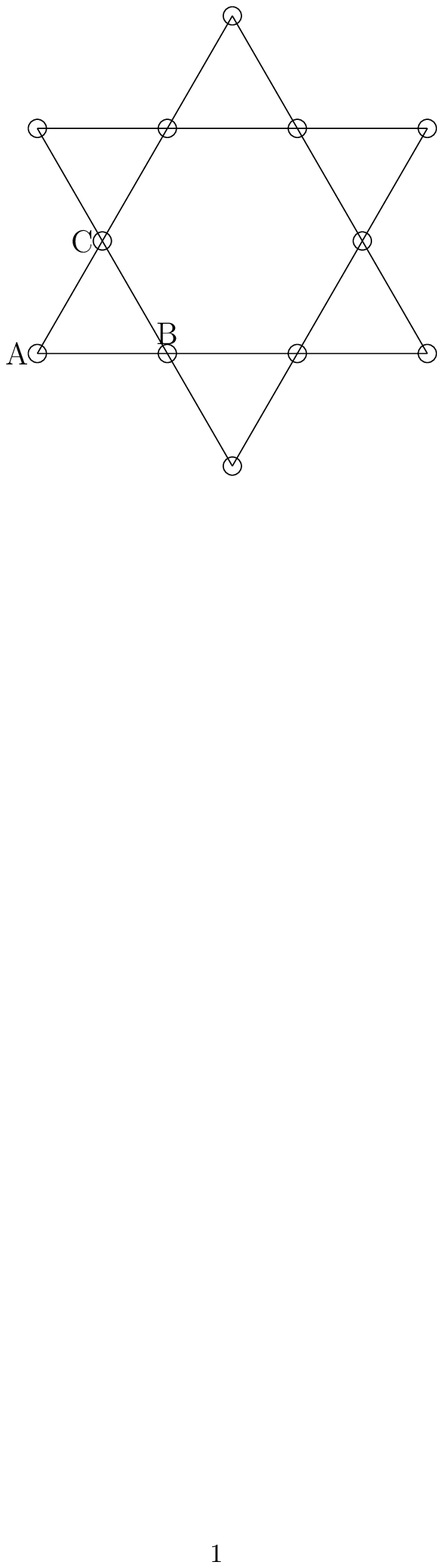}
}\tabf
\subfigure[]{\label{kaglattice}
 \includegraphics[width=4.6cm]{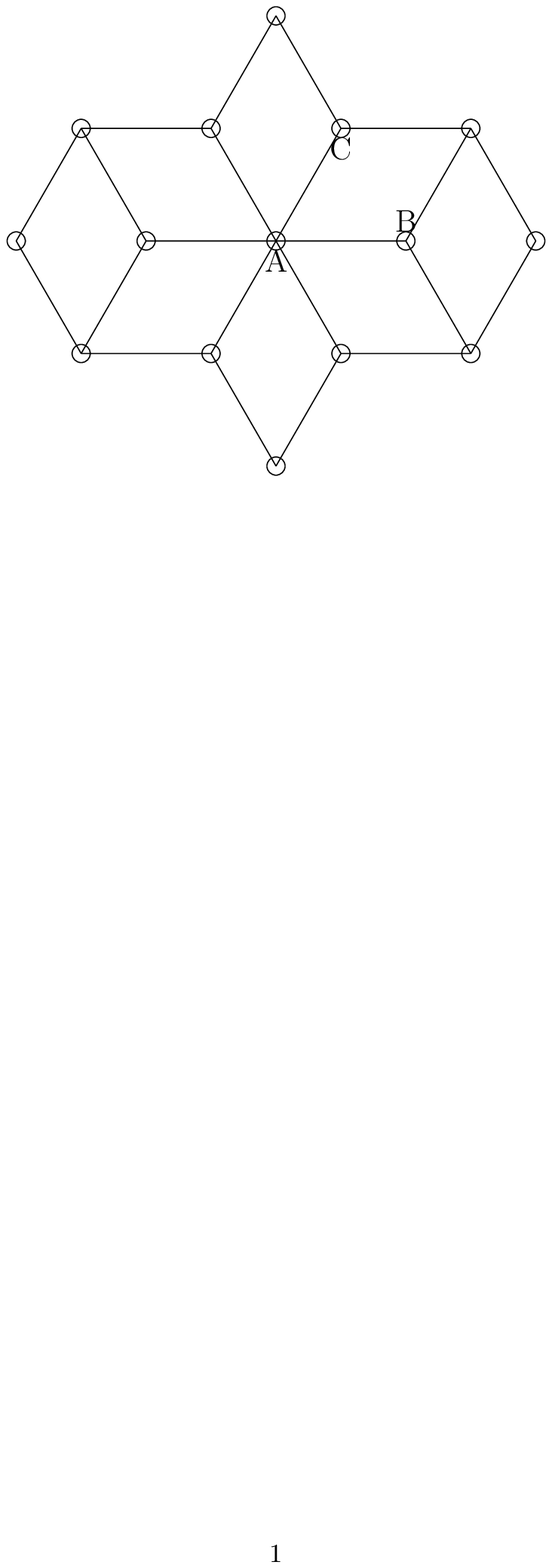}
}
\caption{(a) Diced lattice. (b) Dual to the diced, kagome lattice.}
\label{lattice}
\end{figure}
The diced lattice is the dual to the kagome lattice and is shown in Fig. 1(b) with its three sublattices again labelled as $A,B,C$. The length of segment $AB$ is taken to be one. Here we merely recapitulate the work of Ref. [4] in slightly different form, to correct a very minor error in the density of states. The calculations proceed as in the previous section and the missing details may readily be worked out or obtained from Ref. [4]. The LGF at the origin $\tilde{G}^{\alpha,\alpha}(t, k_x, k_y)$ - where $\alpha = (A, B, C)$, $(k_x,k_y)$ are the Fourier space components as in Section I - for the 3 sublattices $A, B, C$ are given by
\begin{eqnarray}
 & &\tilde{G}^{AA}(t, k_x, k_y) = \frac{4t}{4t^2 - (1+\gamma^2)\omega_k}, \nonumber\\
 & &\tilde{G}^{BB}(t, k_x, k_y) = \frac{4t^2 - \gamma ^2\omega_k}{t[4t^2 - (1+\gamma^2)\omega_k]}, \nonumber\\
 & &\tilde{G}^{CC}(t, k_x, k_y) = \frac{4t^2 - \omega_k}{t[4t^2 - (1+\gamma^2)\omega_k]},
\end{eqnarray}
where $\omega_k \equiv 2\cos{(2ak_x)} + 4\cos{(ak_x)}\cos{(bk_y)}$, $\gamma$ is the $A\leftrightarrow C$ hopping amplitude in units of the $A\leftrightarrow B$ hopping amplitude, which is taken to be $1$. $a = \sqrt{3}/2$ and $b = 1/2$ are lattice distances in units of the inter-atomic distance. Transforming to real space i.e. $\tilde{G} \rightarrow G$, the DOS as a function of the real energy variable $s = \Re{[t]}$ is given by
\begin{eqnarray}
 \rho_{\textrm{diced}}(s) &=& \frac{1}{3\pi}\Im{[\sum_{\alpha}G^{\alpha \alpha}]} \nonumber \\
& & = \frac{\delta(s)}{3} + \frac{4s}{3\pi(1 + \gamma^2)}\Im{[G_{\textrm{tri}}(\tau,0,0)]},
\end{eqnarray}
where $\tau \equiv \frac{4s^2 - 3(1 + \gamma^2)}{2(1 + \gamma^2)}$ and $G_{\textrm{tri}}(t, x, y)$ is the LGF for the triangular lattice; the latter is known from other work\cite{Horiguchi}. Then Eq. (12) simplifies to
\begin{equation}
 \rho_{\textrm{diced}}(s) = \frac{\delta(s)}{3} + \frac{4s}{3(1+\gamma^2)}\rho_{\textrm{tri}}(\tau).
\end{equation}
The DOS for the triangular lattice derived and expressed in Ref. [5] in terms of the complete elliptic functions of the first kind $K(m)$, with $m$ being the parameter, may be rewritten, in the whole range $-1.5 < \tau < 3$, as
\begin{equation*}
 \rho_{\textrm{tri}}(\tau) = \frac{\Re{[K(1 - 1/k(\tau)^2)]}}{\pi^2(2\tau + 3)^{\frac{1}{4}}},
\end{equation*}
where $k(x)^{-2} = \frac{[\sqrt{(2x+3)} - 1]^{3}[\sqrt(2x+3) + 3]}{16\sqrt{(2x+3)}}$.

Here too, it may be checked, say for a particular value of $\gamma = 1$, that $\int_{\frac{-3}{\sqrt{2}}}^{\frac{3}{\sqrt{2}}}\rho_{\textrm{diced}}(s)\,\mathrm{d}s = 1$. In Ref. [4], the Dirac-delta term was not taken into account and the spectral density did not integrate to 1; it is clear that the flat band in the diced lattice must give rise to such a term.
\section{Hyperkagome lattice}
\begin{figure*}[t!]
\centering
\subfigure[ ]{\label{latticeHyper}
\includegraphics*[width=6.1cm]{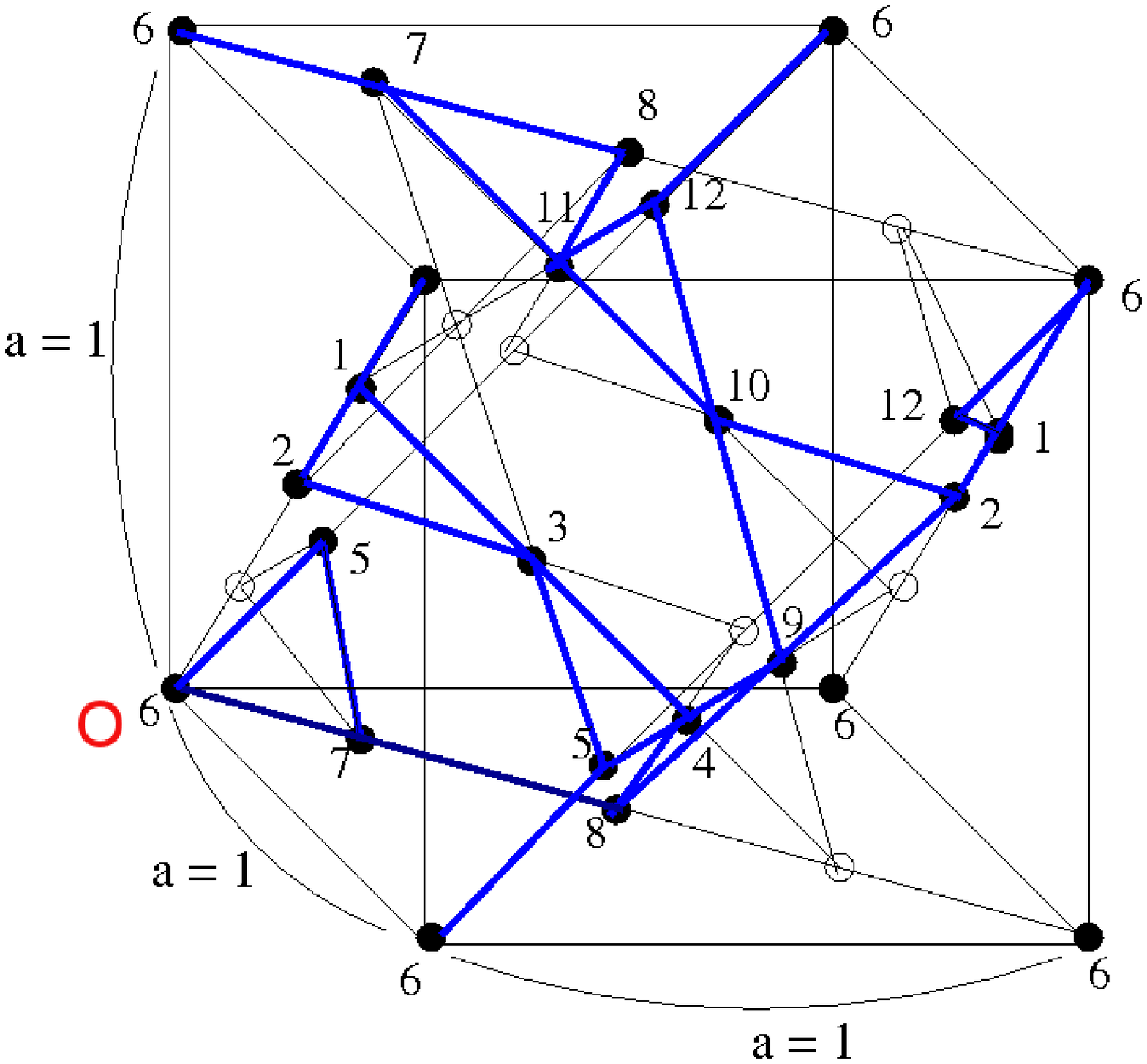}
}\tabf
\subfigure[ ]{\label{latticeHyperDOS}
\includegraphics[width=8.1cm]{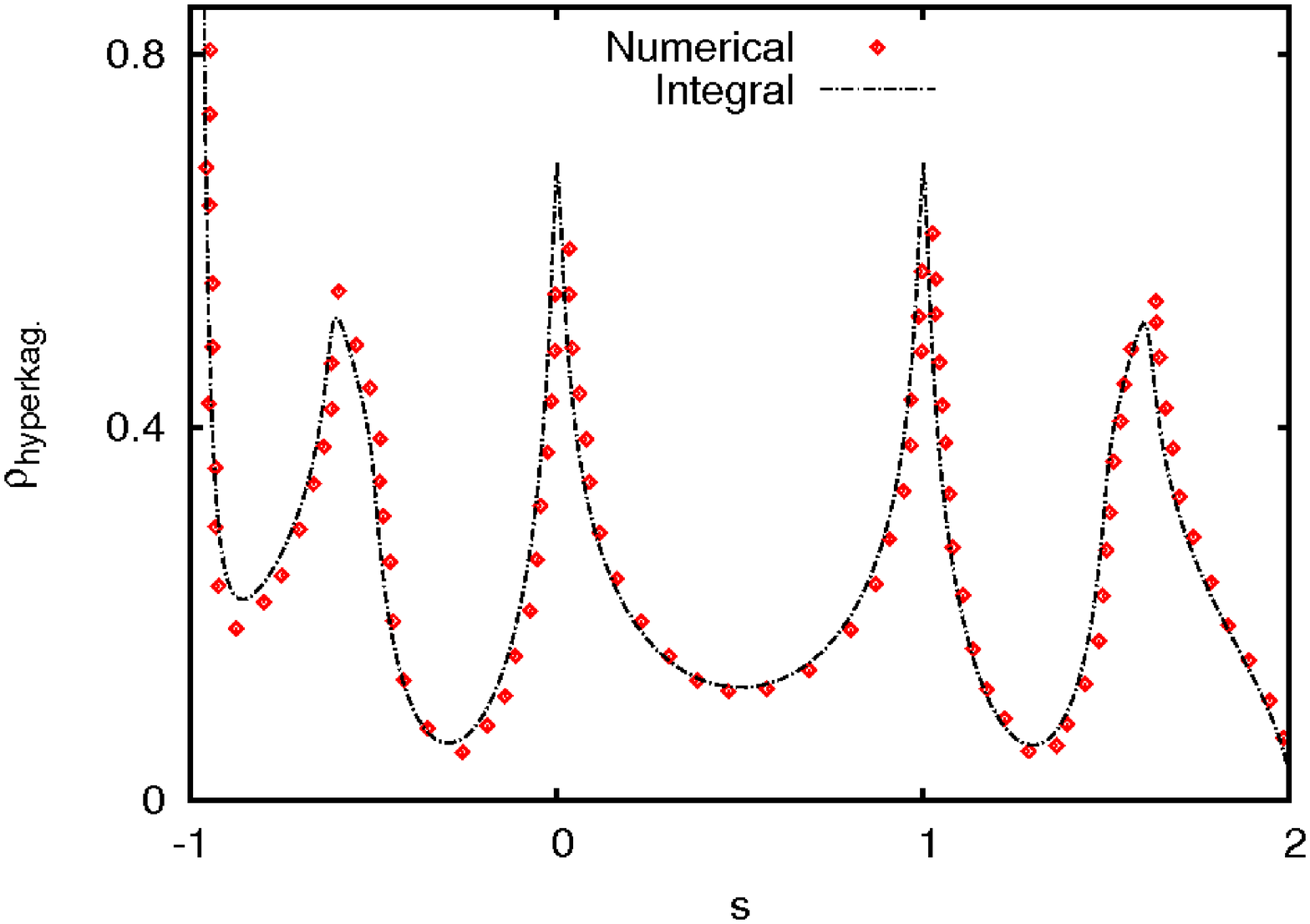}
}
\caption{(a) Unit cell, from Ref. [11] with authors' permission, of the hyperkagome lattice with blue lines and filled circles; the origin is marked with a red $O$. The additional empty circles indicate the underlying pyrochlore lattice. (b) Density of states for the hyperkagome, from evaluating Eq. (16), compared with numerical results from Ref. [9]. Note that the latter had to be multiplied by a factor of 2 to account for their doubling of the energy scales compared to ours.}
\end{figure*}
The hyperkagome lattice can be considered to be a three dimensional version of the kagome lattice. Here, there are corner-sharing tetraheda with one of the atoms in each tetrahedron removed. The unit cell, with 12 atoms, is cubic and is shown in Fig. 2(a). Each side of the cube is taken to have unit length. For example, the lattice point labelled 3 has ($x,y,z$) coordinates ($3/4, 1/4, 1/2$), with respect to the origin $O$, with the nearest neighbours labelled 1, 2, 4, 5. \\
There currently is hardly any exactly known result for this lattice, even for non-interacting models. However it was conjectured\cite{GlasserZucker} that lattices with cubic cells should have a lattice Green's functions expressable in a ``canonical'' integral form, which, in general, can be exactly evaluated; a rationale for this proposition was later given\cite{Glasser}. Glasser's canonical form for the Green's function is 
\begin{equation}
 G = \frac{1}{\pi^3}\iiint_{0}^{\pi}\frac{\textrm{d}x\textrm{d}y\textrm{d}z}{t - F(\cos{x}, \cos{y}, \cos{z})},
\end{equation}
where 
\begin{equation*}
 F(x, y, z) = a_1x + a_2y + a_3z + a_4xy + a_5yz + a_6xz + a_7xyz.
\end{equation*}

Here, based on calculations for the hyperkagome's LGF and DOS, we obtain a form that resembles the canonical form of Glasser. This we solve numerically and compare with other purely numerical results.

Corresponding to each of the 12 lattice points, each of which have 4 nearest neighbours, there will be one LGF equation. These may be obtained in the same way that led Eq. (3) to Eq. (7). The system of equations for the $\alpha$ sublattice, using the corresponding LGF column vector $\tilde{\mathcal{G}}^{\alpha}_{12\mathrm{x}1} \equiv [\tilde{G}^{1,\alpha} \cdots \tilde{G}^{12,\alpha}]^{T}$, is given by
\begin{equation}
 C\tilde{\mathcal{G}}^{\alpha} = 2[\delta_{1,\alpha}\cdots \delta_{12,\alpha}]^{T}_{12\mathrm{x}1},
\end{equation}
where the $12\mathrm{x}12$ coefficient matrix $C$ is given by
\begin{widetext}
\[
 C = \left( \begin{array}{cccccccccccc}
2t & \xi_{-x-z} & \xi_{y-z} & 0 & 0 & \xi_{x+z} & 0 & 0 & 0 & 0 & 0 & \xi_{x-y} \\
\xi_{x+z} & 2t & \xi_{x+y} & 0 & 0 & 0 & 0 & 0 & \xi_{-y-z} & \xi_{-x-y} & 0 & 0 \\
\xi_{-y+z} & \xi_{-x-y} & 2t & \xi_{y-z} & \xi_{x-z} & 0 & 0 & 0 & 0 & 0 & 0 & 0 \\
0 & 0 & \xi_{-y+z} & 2t & \xi_{x-y} & 0 & 0 & \xi_{-x-z} & \xi_{-x+y} & 0 & 0 & 0 \\
0 & 0 & \xi_{-x+z} & \xi_{-x+y} & 2t & \xi_{-y-z} & \xi_{x-z} & 0 & 0 & 0 & 0 & 0 \\
\xi_{-x-z} & 0 & 0 & 0 & \xi_{y+z} & 2t & \xi_{x+y} & 0 & 0 & 0 & 0 & \xi_{-y-z} \\
0 & 0 & 0 & 0 & \xi_{-x+z} & \xi_{-x-y} & 2t & \xi_{x+y} & 0 & 0 & \xi_{y-z} & 0 \\
0 & 0 & 0 & \xi_{x+z} & 0 & 0 & \xi_{-x-y} & 2t & \xi_{y+z} & 0 & \xi_{-x-z} & 0 \\
0 & \xi_{y+z} & 0 & \xi_{x-y} & 0 & 0 & 0 & \xi_{-y-z} & 2t & \xi_{-x+z} & 0 & 0 \\
0 & \xi_{x+y} & 0 & 0 & 0 & 0 & 0 & 0 & \xi_{x-z} & 2t & \xi_{-y+z} & \xi_{-x+z} \\
0 & 0 & 0 & 0 & 0 & 0 & \xi_{-y+z} & \xi_{x+z} & 0 & \xi_{y-z} & 2t & \xi_{-x+y} \\
\xi_{-x+y} & 0 & 0 & 0 & 0 & \xi_{y+z} & 0 & 0 & 0 & \xi_{x-z} & \xi_{x-y} & 2t \end{array} \right),
\]
\end{widetext}
and the symbol $\xi_{\pm \mu \pm \nu} \equiv \mathrm{exp}^{i(\pm k_{\mu} \pm k_{\nu})}$. Now there are 12 systems (for each value of $\alpha$) of 12 linear equations, which is cumbersome to solve by hand as done in previous 2 sections for smaller systems. To obtain the diagonal LGFs, we use Mathematica\cite{Math} to obtain each $\tilde{G}^{\alpha \alpha}$. With these solutions we can use Eq. (6) to express the DOS $\rho _{\mathrm{hyperkag.}}$ for the hyperkagome as
\begin{equation}
 \rho _{\mathrm{hyperkag.}} = \frac{1}{3\pi}\Im{\mathrm{\large{[}}\mathcal{F}^{-1}\{\frac{P(a_0,a_1,a_2,a_3)}{P(b_0,b_1,b_2,b_3)}\}\frac{1}{t+1}\mathrm{\large{]}}}.
\end{equation}
Here the inverse Fourier transform $\mathcal{F}^{-1}$, at $\vec{\mathbf{r}} = \vec{\mathbf{0}}$, of a function $g$ is given by $\mathcal{F}^{-1}g = \frac{1}{\pi^3}\iiint_{0}^{\pi}g(\vec{\mathbf{k}})\textrm{d}\vec{\mathbf{k}}$. The $a_i$ and $b_i$ parameters are only functions of $t$ with no dependence on the lattice momenta i.e.
\begin{widetext}
\begin{eqnarray}
 & &a_0 = 3(-1+12t+72t^2-144t^3-256t^4+448t^5+128t^6-384t^7+128t^8),\nonumber \\
 & &a_1 = -4(1+7t-6t^2-20t^3+16t^4),\nonumber \\
 & &a_2 = 4(1+3t-6t^2), \nonumber \\
 & &a_3 = 12t(1-2t), \nonumber \\
 & &b_0 = 1-16t+112t^2+64t^3-544t^4+256t^5+512t^6-512t^7+128t^8, \nonumber \\
 & &b_1 = -16t(1+t-4t^2+2t^3), \nonumber \\
 & &b_2 = -16t(t-1), \nonumber \\
 & &b_3 = 4(-1+4t-4t^2).
\end{eqnarray}
And the $P$ function is given by
\begin{equation*}
 P(a,b,c,d) = a + b\sum_{i=1}^{3}\cos{(x_{i})} + c\sum_{i=1}^{3}\cos{(x_{i})}\cos{(x_{i+1})} + d\prod_{i=1}^{3}\cos{(x_{i})} + \sum_{i=1}^{3}\cos{(2x_{i})},
\end{equation*}
\end{widetext}
Numerical evaluation of the hyperkagome's DOS\cite{Udagawa} is compared with a numerical evaluation of the integral in Eq. (16) and is shown in Fig. \ref{latticeHyperDOS}. Note that Eq. (16) correctly shows the appearance of a flat band at $t = -1$ as indicated by the simple pole. And as noted from numerical calculations in Ref. [9], there is a drop to zero in the kagome lattice's DOS for the non-interacting Hubbard model (which is merely the second-quantized form of the operator $H(\vec{\mathbf{r}})\,$) but a continuum of energy states in the hyperkagome. We may see this explicitly from Eq. (9) with the appearance of a zero at $2t - 1 = 0$; no such apparent zero occurs for the case of the hyperkagome in Eq. (16), verified by numerical integration of the same as shown in Fig. \ref{latticeHyperDOS}. A continuum of states is, of course, indicative of conductive behaviour.\\
Many similar integrals - which can be exactly solved - appear in the literature\cite{GlasserZucker, Glasser1} for the lattices with a cubic unit cell but not with all the sums and products of $H$ appearing together. In particular, the last summations of $\cos{2x}$ in the $P$-function do not appear in the canonical form in Eq. (15),(16). It is unclear if these terms might suggest a generalization of the canonical form or if integrability is destroyed by their presence. We have currently found no way to exactly solve the Eq. (16).

To summarize, we have calculated lattice Green's functions and density of states for three related lattices: the kagome, diced and hyperkagome. For the hyperkagome lattice, however, we are able to reduce the obtained integral to a form approximately like Glasser's canonical form but with some additional terms that appear to destroy integrability. Due to the relatively succinct functional form obtained for the integral and the numerical correspondence of this result with other purely numerical evaluations for the density of states, we believe that it might be plausible to obtain a closed form solution. Exact expressions for lattice density of states ought to be useful in dynamical mean field theory calculations\cite{Lee}.\\
\acknowledgments

We thank M. Udagawa, B. H. Bernhard and S. G\"{u}rtler for useful discussions. VKV is partially supported by the Bonn-Cologne Graduate School within the Deutsche Forschungsgemeinschaft.
\bibliographystyle{unsrt}
\bibliography{Ref}

\end{document}